\documentclass{article}

\usepackage{arxiv}

\usepackage[utf8]{inputenc} 
\usepackage[T1]{fontenc}    
\usepackage{hyperref}       
\usepackage{url}            
\usepackage{booktabs}       
\usepackage{amsfonts}       
\usepackage{nicefrac}       
\usepackage{microtype}      
\usepackage{lipsum}
\usepackage{fancyhdr}       
\usepackage{graphicx}       
\graphicspath{{media/}}     

\title{
 
 A Self-Improvable Polymer Discovery Framework Based on Conditional Generative Model}

\author{
  Arash Khajeh\textsuperscript{1†*}, Xiangyun Lei\textsuperscript{†1}, Weike Ye\textsuperscript{†1}, Zhenze Yang\textsuperscript{1,2}, Daniel Schweigert\textsuperscript{1}, Ha-Kyung Kwon\textsuperscript{1}\\
  \textsuperscript{1}Toyota Research Institute, \\
  4440 El Camino Real, Los Altos, California 94022, United States of America.\\
  \textsuperscript{2}Department of Materials Science and Engineering, \\
  Massachusetts Institute of Technology, \\
  77 Massachusetts Ave., Cambridge, Massachusetts 02139, United States of America \\
  \texttt{arash.khajeh@tri.global} \\
}

\DeclareUnicodeCharacter{2212}{-}

\begin{document}
\maketitle

\begin{abstract}
In this work, we introduce a polymer discovery platform to efficiently design polymers with tailored properties, exemplified by the discovery of high-performance polymer electrolytes. The platform integrates three core components: a conditioned generative model, a computational evaluation module, and a feedback mechanism, creating a self-improving system for material innovation. To demonstrate the efficacy of this platform, it is used to design polymer electrolyte materials with high ionic conductivity. A simple conditional generative model, based on the minGPT architecture, can effectively generate candidate polymers that exhibit a mean ionic conductivity that is significantly greater than those in the original training set. This approach, coupled with molecular dynamics simulations (MD) for testing and a specifically planned acquisition mechanism, allows the platform to refine its output iteratively. Notably, after the first iteration, we observed an increase in both the mean and the lower bound of the ionic conductivity of the new polymer candidates. The platform's effectiveness is underscored by the identification of 14 polymer repeating units, each displaying a computed ionic conductivity surpassing that of Polyethylene Oxide (PEO). The performance of these polymers in MD simulations verifies the platform's efficacy in generating potential polymer candidate materials. Acknowledging current limitations, future work will focus on enhancing modeling techniques, evaluation processes, and acquisition strategies, aiming for broader applicability in polymer science and machine learning.
\end{abstract}


\section{Introduction}

The domain of polymer science is pivotal in shaping advancements across diverse technological arenas. Polymers, with their extraordinary adaptability and customizability, cater to a wide range of applications, spanning from biodegradable materials and high-performance aerospace composites to conducting elements in electronic devices and smart materials in sensor technologies. Notably, polymer electrolytes play a crucial role in the field of energy storage, exemplifying the versatility of polymers \cite{PolymerElectrolytes1, PolymerElectrolytes2, PolymerElectrolytes3, PolymerElectrolytes4}. Currently, several challenges are associated with liquid electrolytes, which are the commercially used materials in Li-ion batteries, including flammability~\cite{han2023incombustible, https://doi.org/10.1002/adfm.202008644}, toxicity~\cite{lebedeva2016considerations, strehlau2017towards}, and instability of the electrode-electrolyte interface due to lithium dendrite formation~\cite{xiao2019lithium, takeda2016lithium}. These issues, along with the several advantages of polymer electrolytes, such as better adaptability to current manufacturing processes compared to ceramics~\cite{yu2021review} and lower cost compared to ionic liquids~\cite{ma2023ionic}, make polymer electrolytes promising candidates for future generations of batteries and energy storage devices. These diverse benefits highlight the transformative potential of polymer electrolytes in revolutionizing energy storage technologies, aligning with broader advancements in polymer science for sustainable and high-performance applications.

Recent advancements in the ionic conductivity of polymer electrolytes, crucial for energy storage systems, have been driven by innovative strategies in polymer structure manipulation. These include crosslinking, blending with additives~\cite{Tang2012, Bandara1998}, and copolymerization~\cite{maia2022designing, bouchet2013single, C3CC49588D}. For instance, He et al.~\cite{he2020high} enhanced Li-ion mobility and electrochemical stability by altering the structure of solid polymer electrolytes, moving the carbonate group to the side chain and using a hydrocarbon backbone. This change stabilized the polycarbonate backbone and achieved a conductivity of about 1.1 mS/cm at room temperature with a plasticizer\cite{he2020high}. Similarly, Zhang et al.\cite{zhang2019cross} improved the amorphous domain of polyethylene oxide (PEO) through crosslinking with tetraglyme (TEGDME) and the integration of a rigid linear oligomer, tetraethylene glycol dimethacrylate (TEGDMA), resulting in an electrolyte system that demonstrated an ionic conductivity of $2.7 \times 10^{-4}$S/cm at room temperature. This system also featured a favorable transference number and low interfacial resistance, showing high potential for energy storage applications. In another approach, Sun et al.\cite{sun2020fast} developed lithium polysulfides-grafted PEO-based electrolytes via in-situ reduction of sulfur copolymers with poly(ethylene glycol) methyl ether methacrylate (PEGMA). This technique enabled fast Li\textsuperscript{+} transport and a close electrode-electrolyte interface, achieving dendrite-free Li-metal deposition and maintaining cell capacity over 1200 cycles at 1 C with only a 0.024\% capacity decay per cycle. Additionally, Lin et al.\cite{lin2021block} designed novel block copolymer electrolytes with three-dimensional networks, enhancing both ionic conductivity to $5.7 \times 10^{-4}$S/cm and lithium ion transference number to approximately 0.49. They also expanded the electrochemical window to about 4.65V (vs. Li\textsuperscript{+}/Li) and improved mechanical strength at 55~°C by optimizing the ratio of functional units in the copolymers. Despite these significant advances, achieving ion transport properties in solid polymer electrolytes that match those of liquid counterparts remains a formidable challenge, underscoring the ongoing need for innovation in this field.

Identifying polymers with the optimal blend of properties for specific applications, including polymer electrolytes for energy storage, is a significant scientific challenge. The complexity of polymer structures, combined with the necessity to balance multiple properties like mechanical strength, electrical conductivity, and thermal stability, makes the discovery process highly intricate. Traditional methods, though foundational, are often limited in their ability to rapidly identify innovative materials, including advanced polymer electrolytes for next-generation energy storage solutions. This limitation points to the need for more efficient and comprehensive approaches to polymer discovery.

Machine learning, particularly in the realm of generative modeling, presents a transformative approach to this challenge. Generative models in machine learning have shown promise in various domains, including material science, by enabling the exploration of vast chemical spaces with unprecedented efficiency. These models can quickly navigate the intricacies of polymer chemistry, suggesting novel and plausible compositions and structures for investigation, thereby streamlining the discovery process.

Within this realm, conditioned generative modeling presents a particularly relevant technique. By training models on specific conditions or properties, it becomes possible to generate content that meets predetermined criteria. In the current landscape, while conditioned generative models specifically for polymers are still emerging, the concept of integrating machine learning with material science to tailor polymer properties is gaining traction. Our work contributes to this field by introducing a comprehensive polymer discovery platform that leverages the principles of conditioned generative modeling. This platform is not limited to merely suggesting potential polymer candidates but is designed to iteratively improve and refine its suggestions based on continuous feedback and evaluation. Such a self-improving system embodies a significant leap from traditional methods, offering a more holistic and efficient pathway to polymer material innovation \cite{PolyG2G,P1M,KIM2021110067}.

 In this paper demonstrates the application of our discovery platform in the realm of polymer electrolytes for energy storage technologies. We focus on polymer electrolytes due to their crucial role in the efficiency and safety of devices such as lithium-ion batteries and supercapacitors. We limit the scope of this study to dry (solid) linear chain homopolymers to utilize the HTP-MD~\cite{xie2022htpmd}, a recently developed large database of polymer electrolyte properties computed from MD simulations. While we explored the generation process and compared the performance of different model architectures for the inverse design of polymer electrolytes in our concurrent study~\cite{Yang2023}, this study focuses primarily on integrating the necessary components to propose a discovery framework and evaluate its outcomes. This proof-of-concept study aims to demonstrate that, starting from a set of polymers with computed properties via MD simulations, our framework can design polymers that compete with Polyethylene Oxide (PEO), one of the best-known polymers in dry form in terms of ionic conductivity. We show that our platform successfully designs polymer electrolytes with ion conductivities superior to the current benchmark, PEO, as assessed by molecular dynamics (MD) simulations.

While many other requirements, such as synthesizability, thermochemical stability, compatibility with other components, and in-device performance of newly generated polymer electrolytes, should be considered in the future, the outcome of this study is a small step toward efficiently navigating the large design space. We believe the results of the current study highlight the potential of our platform not only in accelerating the discovery of high-performance polymer electrolytes but also in setting a precedent for its application across various polymer-based technologies.

\section{Platform}
\begin{figure}[h]
    \centering
    \includegraphics[width=0.7\textwidth]{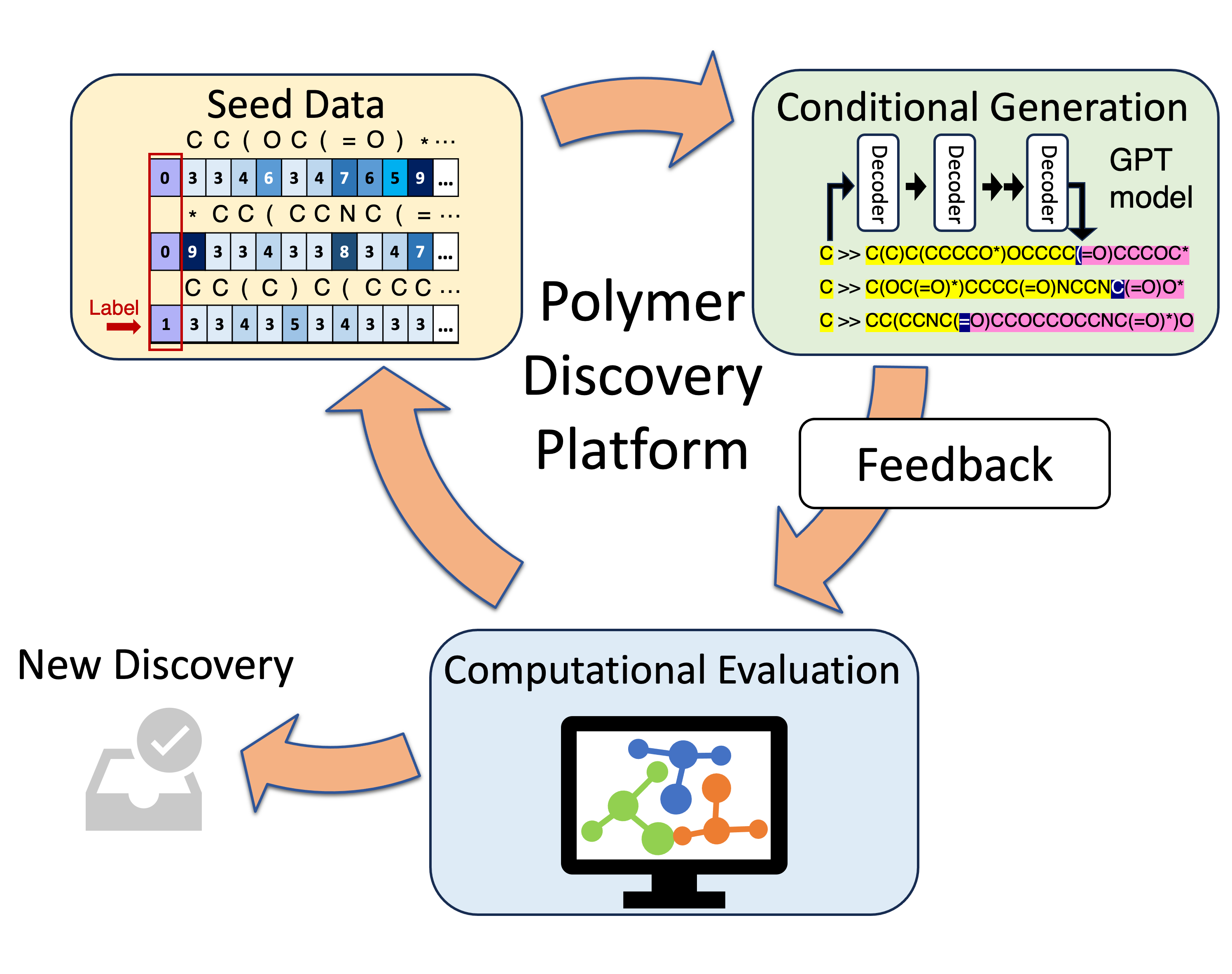}
    \caption{Schematic illustration of the platform. }
    \label{fig:Schema}
\end{figure}
Here, we present a platform structured around an iterative, self-sustaining, and self-improving workflow, comprising three essential components: a conditional generative model, a computational evaluation module, and a feedback mechanism. This integrated system allows for continuous refinement and evolution of the discovery process, which we term a "discovery campaign."

The conditioned generative model, the heart of this platform, is tasked with proposing potential polymer candidates. Tailored to incorporate specific target properties, this module is responsible for generating polymers, either by constructing repeating units, oligomers, polymer chains, or 3D structures. For this proof of concept, our focus is on generating the 2D representation of repeat units of polymers. Several aspects heavily influence the performance of the generative model: the (seed) data, the model architecture and hyperparameters, and the training strategies. Different from traditional regression models where a numerical loss is clearly defined, the generative tasks are more ambiguous to evaluate and often require domain knowledge. The process of formulating a comprehensive and domain-relevant evaluation schema and performing the benchmarking across a series of model architectures and training strategies presents its own set of challenges. Our separate research paper~\cite{Yang2023} addresses these challenges, offering a systematic approach for benchmarking different generative models in different tasks and proposing a comprehensive set of evaluation metrics for the task of generating alternative polymers. 

Once a batch of polymers is proposed by the generative model, the evaluation module takes over. This component is responsible for assessing the target properties of the proposed polymers, employing both simulation and experimental validations. For computational evaluation, in the current study, we rely on molecular dynamics (MD) simulations (see Experiment Setup for details). Experimental validation serves as the definitive confirmation for our newly discovered polymers. While this crucial phase falls beyond the scope of the current study, we intend to conduct experimental tests on these candidates in a subsequent phase of our research. 

Establishing a feedback mechanism is pivotal to allowing active learning and continuous self-improvement of the model. At the end of each campaign iteration, all computed results are recorded in a database, and strategically sampled for enriching the training data (please see the details of sampling in~\ref{Experimental Setup}). The model is then retrained to the new data to become increasingly adept at targeting the desired polymers. For future improvements, the uncertainty of the generative model ought to be quantified, and sophisticated acquisition strategies to balance exploration and exploitation are to be tested and implemented.

The initialization and deployment of the platform are crucial steps of the discovery campaign. The quality and distribution of the seed data are paramount to the campaign's success, and they represent the main responsibility of the user. With the appropriate seed data and clearly defined target properties, the platform is designed to be capable of operating autonomously, minimizing the need for further human intervention. This autonomous nature highlights the platform's potential for high-throughput, efficient discovery campaigns and opening new horizons in the field of polymer science.

\section{Experiment Setup} \label{Experimental Setup}

\subsection{Scientific task}
To demonstrate the usage of the platform, we deployed it to identify polymer electrolytes with high ionic conductivity, as advancements in energy storage technologies are increasingly contingent on the development of such materials. Despite significant progress, finding materials that surpass the performance benchmarks set by current standards, such as Polyethylene Oxide (PEO), remains a challenge. PEO, while widely used, falls short of meeting the growing demands for efficiency and stability under varied operational conditions. This limitation is particularly pronounced in applications requiring high charge-discharge cycles and under extreme temperature variations.  Our endeavor in the current study mainly aims to demonstrate the power of this discovery framework for identifying polymer electrolytes with superior ionic conductivities. While this study does not specifically target thermal stability or varied operational conditions, the framework has the potential to be extended in future work to address these critical challenges, potentially leading to safer, more durable, and higher-capacity energy storage solutions.

\subsection{Conditional generation}

The core of this demonstration revolves around a conditional generative model based on the minGPT~\cite{mingptrepo} architecture. When trained, this generative model is capable of generating a polymer electrolyte candidate given a lead prompt signaling what kind of property is needed (with high or low ionic conductivity), just like the popular large language models. The specific model used utilizes the gpt-nano architecture with approximately 120,000 trainable parameters, and we used the Simplified Molecular Input Line Entry System (SMILES)~\cite{toropov2005simplified} code of polymers' repeating units to represent the polymer electrolyte candidates. Although relatively simple in its design, it is powerful in its application. To direct the model towards polymers with high ionic conductivity, we implemented a method to incorporate the properties of the input during training. This involved the modification of tokenized SMILES strings of known polymer electrolytes prefixed by their ionic conductivity classes. Specifically, given the range (0.007-0.506 mS/cm) and distribution (mean= 0.062 mS/cm, std= 0.036 mS/cm) of ionic conductivity in our dataset, we assigned different class labels to high-conductive (top 5\%) and low-conductive (lower 95\%) polymers and used this class as the leading digit in the input to the model. Therefore, the dividing line between the "high-conductivity" and "low-conductivity" categories is set to be 0.012 mS/cm, and this is fixed throughout the experiment. For example, a polymer with an ionic conductivity of 0.18 (mS/cm) is labeled with class 1 (high conductivity), while another polymer with an ionic conductivity of 0.06 (mS/cm) is labeled with 0 (low conductivity). As the data set evolves over time, the allocation of polymers to the respective high- and low-conductive groups will change accordingly. Additionally, to maintain the importance of the property class in comparison to the lengthy SMILES, and to ensure the model can be effectively guided towards desirable structures, we replicate the property class five times, converting the desired class tokens to "11111", which is shown to result in better polymer electrolyte candidate empirically. Therefore, the effective prompt that signals the conditioned generative model to generate polymer electrolyte candidates with high ionic conductivity is simply a leading string of "11111", and the model will then complete the string by generating a SMILES code that represents a specific polymer candidate.

Inspired by the simple design of PEO with a very short repeat unit  (OCC) and high ionic conductivity (1.15 mS/cm, at 353 K, $Li^+.TFSI^-$ molality= 1.5 mol/kg),  during the iterative polymer generation loops, the model is also biased towards generating small repeating units with SMILES strings containing 10 or fewer tokens. This approach, albeit unrefined, proved crucial for the model's ability to generate high-conductivity polymers. The reason for the effectiveness of conditioning on short repeat units can be the short distance between negatively charged atoms, such as oxygen atoms, in the polymer backbone that coordinates with Li ions. An effective coordination environment can help with salt dissociation to individual ions and the easier hopping of cations from one coordination site to another~\cite{roy2016correlation, zhang2023bimetallic}. Regardless of the imposed restriction on the length of repeat units (10 and fewer characters), the model itself shows a tendency toward generating short repeat units, which is mainly due to oversampling PEO that will be discussed in Section~\ref{Data}.  

\subsection{Model details} \label{model}
The generative model we implemented here is based on a minimal implementation of the GPT model.~\cite{mingptrepo}. The model first converts a sequence of tokens (SMILES vocabulary) to two embeddings, including token embedding and positional embedding. Token embedding represents the meaning of individual words or symbols in a high-dimensional space, while positional embedding encodes the order or position of tokens within a sequence to provide contextual information to the model. These embeddings are then passed through multiple layers of transformer blocks. Each transformer block mainly consists of a multi-head masked self-attention layer and a feed-forward neural network, following the original transformer architecture~\cite{NIPS2017_3f5ee243}. The loss function is cross entropy loss comparing tokens in the generated and actual sequences. 

We performed a grid search to tune the hyperparameters of minGPT and optimize the performance of the model. The mean values of six metrics (chemical validity, uniqueness, novelty, synthesizability, similarity, diversity) are utilized as the evaluation metric for the model's performance. It is important to note that in this study, novelty is defined as a polymer structure that has not been seen by the model during the training process and is not based on literature novelty. We used 3 independent hyperparameters, including "training epoch," "model architecture," and "temperature." The "training epoch" varies from 1000 to 10000 with a uniform interval of 1000. The "model architecture" corresponds to HuggingFace’s Transformer-based model architecture.~\cite{Wolf2019} We select 3 different architectures which are "gpt-2", "gpt-mini" and "gpt-nano". In terms of the "temperature", the values are from 0.1, 0.5 and 1.0. A higher temperature results in higher "creativity" for the GPT model. For the optimal performance, the "training epoch" is 6000, the "model architecture" is "gpt-nano" and the "temperature" is 1.0. As a result, the number of attention heads in the minGPT model is 3, and the hidden dimensions are set to 48, which results in around 0.12M parameters. The hyperparameter selection and the grid search results for the GPT model can be in the Supporting Information Table. S1, S2, S3, and S4. For more details, please refer to our concurrent work~\cite{Yang2023}.

\subsection{Data} \label{Data}

The dataset used in this study is a subset of polymers from High-Throughput Polymer Design - Molecular Dynamics (HTP-MD) database~\cite{htpmdwebapp, xie2022htpmd}, consisting of 6024 linear chain homopolymers. Selected polymers were all unique and composed of H, C, F, S, P, O, N, elements, previously filtered from 53362 structures in the Zinc database~\cite{doi:10.1021/ci049714+} to ensure both synthesizability and potential application as electrolytes~\cite{xie2022accelerating}. The ionic conductivity values for polymer-Li$^+$.TFSI$^-$ systems computed from MD simulations performed in the large atomic molecular massively parallel simulator (LAMMPS)~\cite{PLIMPTON19951} with the interaction parameters from Polymer Consistent Force Field (PCFF$^+$)~\cite{https://doi.org/10.1002/jcc.540150708, https://doi.org/10.1002/(SICI)1097-0126(199711)44:3<311::AID-PI880>3.0.CO;2-H}. The PCFF+ force field has been employed in previous studies to explore various properties of electrolyte systems~\cite{xie2022accelerating, PhysRevLett.122.136001, france2019effect, wang2020toward, wahlers2016solvation, meng2019superior, feng2022room}. Additionally, multiple studies have compared the molecular dynamics (MD) simulation results obtained using the PCFF+ force field with experimental data and density functional theory (DFT) calculations for both polymer~\cite{xie2022accelerating, doi:10.1021/acs.chemmater.8b01955, doi:10.1021/acs.macromol.1c01028} and liquid~\cite{crabb2020importance, rozanska2015automatic, D2CP05423J} electrolytes. The simulations carried out for dataset generation have been previously performed in another study~\cite{xie2022htpmd}, at 353 K and the salt molality of 1.5 mol/kg. More details on the details of simulations and calculation of ionic conductivity can be found in the original work~\cite{xie2022htpmd,githubcode}. 

To skew the model towards high-conductivity polymers, we randomly oversampled the top 5\% of polymers in terms of ionic conductivity. This provides a train set that includes the same number of polymers from low-conductive and high-conductive classes. Additionally, PEO was added to the train set and oversampled 4000 times in the seed data. This method of selective oversampling was shown to be instrumental in guiding the model towards generating more promising polymer candidates. It should be noted that in the current study, oversampling was performed by including duplicate SMILES strings, and we have not tried other methods, such as randomization~\cite{arus2019randomized}.  

\subsection{Evaluation and feedback}
For evaluation purposes, each iteration of the discovery process involved the generation of 50 polymer candidates, with their ionic conductivities evaluated through molecular dynamics simulations. These simulations adhered to the same protocol used in creating a previous dataset (HTP-MD:~\cite{htpmdwebapp}). 

We carried out molecular dynamics (MD) simulations on polymer-(Li$^+$.TFSI$^-$) systems at a temperature of 353 K and a salt concentration of 1.5 mol/kg using LAMMPS~\cite{PLIMPTON19951} with the Polymer Consistent Forcefield (PCFF+)~\cite{https://doi.org/10.1002/jcc.540150708}. The simulation process included initial relaxation and equilibration of the polymer-salt mixture, followed by a production phase to gather data for computing ion transport properties, such as ionic conductivity. The equilibration stage involved running sequential NVT (constant number of particles, volume, and temperature) and NPT (constant number of particles, pressure, and temperature) ensembles to achieve densities close to theoretical values. For the production run, an NVT simulation at 353 K was conducted for 5 ns with a 2.0 fs time step. The resulting trajectories were then analyzed to compute ion transport properties using the cluster Nernst-Einstein equation~\cite{PhysRevLett.122.136001}. The analysis code used to compute ionic conductivity is consistent with the one used to generate the HTP-MD database and is available at \href{https://github.com/TRI-AMDD/htp_md}{https://github.com/TRI-AMDD/htp$\_$md}. More details about MD simulations, dataset composition, and computing ionic conductivity have been included in ~\ref{Data} section, as well as previous studies~\cite{xie2022accelerating, xie2022htpmd, githubcode, htpmdwebapp}. The above-described simulation protocol has been previously employed to assess the computed ion transport properties of polymers in the HTP-MD database and compare them with the experimentally measured ionic conductivity values.~\cite{xie2022accelerating}  

The repeat units of the generated polymers are polymerized to have at least 150 heavy atoms (non-H) in their backbone, with the two ends terminated by methyl groups. This approach is consistent with the method used to generate the HTP-MD database, allowing us to compare the performance of newly generated polymers with the training set. Also, to ensure robustness, each candidate underwent five independent simulation replicas to determine its conductivity. Given the randomness in MD simulation results originating from different conformation sampling, this rigorous step was crucial for ascertaining the potential of each proposed polymer.

The feedback mechanism of our platform plays a vital role in its iterative learning process. After evaluation, we added both PEO and newly discovered polymers showing conductivity higher than PEO to the train set and oversampled 4000 in total from all newly added polymers. This is to ensure the model can still explore polymers that are different from PEO. This enriched dataset was then used to retrain the generative model, thereby enhancing its ability to propose increasingly relevant and high-performance polymers in subsequent iterations. For demonstration purpose, the discovery campaign is conducted for two iterations.

\section{Results}

\subsection{Conditional Generation}
\begin{figure}[h]
    \centering
    \includegraphics[width=0.7\textwidth]{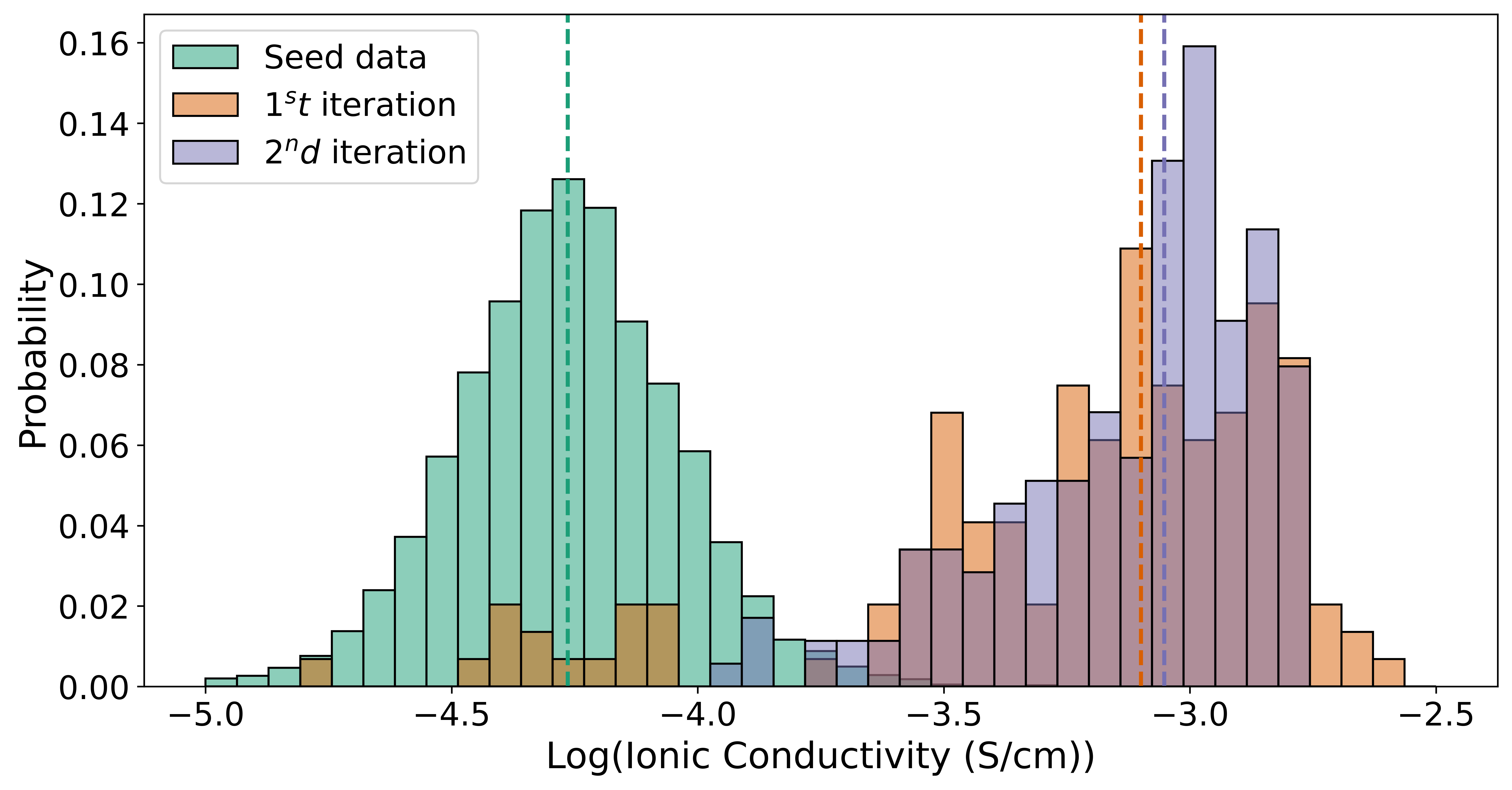}
    \caption{Comparison between the distribution of ionic conductivity in the train set (green)  generated set in the first iteration (orange), and second iterations (purple). The dotted vertical lines show the mean ionic conductivity in each distribution. The values in the histograms are computed from MD simulations.}
    \label{fig:comparison between distributions}
\end{figure}
Generative models have gained substantial traction within the scientific community, showcasing their effectiveness and widespread adoption. However, within the realm of materials science, the primary challenge often lies not in simply generating physically reasonable candidates but rather in producing materials with specifically targeted properties, a process commonly referred to as inverse design. To illustrate this point, consider the case of polymer electrolytes, where researchers aspire to discover novel polymers exhibiting enhanced ion conductivity. Historically, experimental development of new polymer electrolytes that can compete with PEO has been mainly focused on different trial errors, including altering the O/C ratio in the polymer backbone~\cite{halat2021modifying, snyder2021improved, fang2023ion}. However, this approach is inefficient, and the previous attempts were not completely successful in finding polymers that can replace the liquid electrolytes. 

In this study, we present an approach that leverages the transformer-based generative model with a modified encoding approach to achieve conditional generation based on the desired property (see Experiment Setup for details). The encoding contains two layers of information: the SMILES representation of the repeating unit and the class of the ion conductivity. When trained with more than 6000 data, the model not only learned how to generate chemically valid SMILES string as polymer repeating units but also the relationship between the repeating unit and the ion conductivity  (please also see~\cite{Yang2023}). The outcome of the model is a generative batch of polymer repeating units, which exhibits a shifted distribution of ion conductivity with a notably higher mean value (0.75 mS/cm) when compared to the training set (0.06 mS/cm, excluding the added PEO polymers) (Figure 2). It is a remarkable improvement by a factor of 12.

\subsection{Active learning}

While the initial success in conditional generation is commendable, we recognize that the current data on polymers remains sparse. To foster creativity and generate a wider variety of candidates, it's crucial to expose our model to new samples continuously. This involves integrating a feedback loop from the verification module in our platform.

Presently, our method involves using Molecular Dynamics (MD) simulations to estimate ionic conductivity in newly generated polymers. Polymers demonstrating higher ion conductivity are then added to our training dataset. We examined how the distribution of polymers evolved across various iterations of active learning. The immediate finding is that the average ion conductivity of polymers generated by a model after one retraining cycle exceeded that of the initial model by 15\%, as shown in the box-and-whisker plot in Figure 3. Further, the lowest ion conductivity reported from the 2nd iteration batch is 3.6 times that of the 1st iteration. Furthermore, in Figure 4, we highlight new repeating units discovered through this process, which show higher ion conductivity compared to those in the original training set. This progression indicates the model's potential for ongoing enhancement, leading to increasingly effective outputs via a systematic feedback approach. Other ion transport properties of the generated polymer in the two iterations, including density, ion diffusivity, ionic conductivity, and transference number, are listed in Supporting Information (Table.S5 and S6). 

Although we observed an increase in the minimum and average ionic conductivity, the improvement in ionic conductivity and the number of polymers outperformed the baseline levels (7 polymers in each iteration), and there was, in fact, a decrease in the maximum ionic conductivity of generated polymers in the second iteration. We believe this decrease is due to the exhaustion of the limited search space. Since we are exploring linear chain homopolymers composed of only a few heavy atoms, the search space is extremely narrow. The limited search space is also the reason that only two iterations of exploration are performed. We believe that by extending the chemistry to a wider range of atom types and incorporating more diverse polymer structures (e.g., aromatic structures), a saturation of performance increase would occur at later stages. This highlights the necessity of developing more creative generative models that can extrapolate the search space, which has been the focus of other researchers' studies \cite{PolyG2G}.

\subsection{Discovered polymers}

In Figure~\ref{fig:discovered-polymers}, we introduce 14 generated polymer repeating units whose ionic conductivities, as confirmed by MD simulations, surpass that of PEO. It is important to reiterate that PEO currently holds the record for one of the highest ion conductivity in the form of dry (solid) polymers with around 1 mS/cm conductivity at 353 K, and $Li^+.TFSI^-$ molality of 1.5 mol/kg) and is still considered a baseline material in this field.

Among these polymer repeat units, several polyacetals stand out. Polyacetals are polymers with a high oxygen-to-carbon ratio, similar to PEO, which facilitates efficient lithium salt solvation and creates effective pathways for lithium ion transport. For example, poly(1,3-dioxolane) (P(EO-MO), *OCCOC*), a polyacetal with a repeating unit of 1,3-dioxolane, demonstrates considerable promise. Its MD-calculated ion conductivity is 1.515 (± 0.199) mS/cm. Although experimental measurements show lower conductivity for P(EO-MO) at 0.4 mS/cm, its potential as a polymer electrolyte candidate remains significant due to its improved ion transport efficacy~\cite{snyder2021improved}. Similarly, poly(diethylene oxide-alt-oxymethylene) (P(2EO-MO)), which appeared as *OCCOCCOC* in the first iteration, has been previously synthesized and investigated, showing slightly lower ionic conductivity of 1.1 mS/cm compared to PEO's 1.5 mS/cm at 90°C~\cite{doi:10.1021/acs.macromol.7b02706}. Another notable polymer is polyethylene oxide-alt-trimethylene oxide (P(EO-TMO)), which appeared as *OCCOCCC* in the second iteration. Previous molecular dynamics simulations have supported the higher ionic conductivity of P(EO-TMO) compared to PEO~\cite{doi:10.1021/acs.macromol.5b01437}, but we couldn't find any experimentally measured ionic conductivity for this polymer. 

Our research uncovered that many of the candidate polymers feature new elements like nitrogen and sulfur, marking a shift from the conventional focus on polycarbonates composed solely of carbon and oxygen. Notably, the polymer with the highest conductivity discovered in our study contains the repeating unit \textit{ONCCOC}. Molecular dynamics simulations demonstrated that this polymer exhibits an average ionic conductivity roughly twice that of polyethylene oxide (PEO). Despite the absence of previous experimental or theoretical studies on this specific polymer, related research on polyethylenimine (PEI, CCN) polymer electrolytes~\cite{BAYRAKPEHLIVAN2014164, yang2020robust} and their blends with PEO~\cite{pritam2020selection, lehmann2020well} has been documented. However, many of the generated polymers include bonds such as N-O or S-O and motifs like O-O-NH or O-NH-O, which present significant synthetic challenges due to their physically unrealistic structures. Nonetheless, the success of our framework in consistently generating polymers with improved ionic conductivity in MD simulations and the appearance of many of the generated candidates in the literature highlights the innovative potential of our models and greatly expands the possibilities for future research in this field.

\begin{figure}[h]
    \centering
    \includegraphics[width=1.0\textwidth]{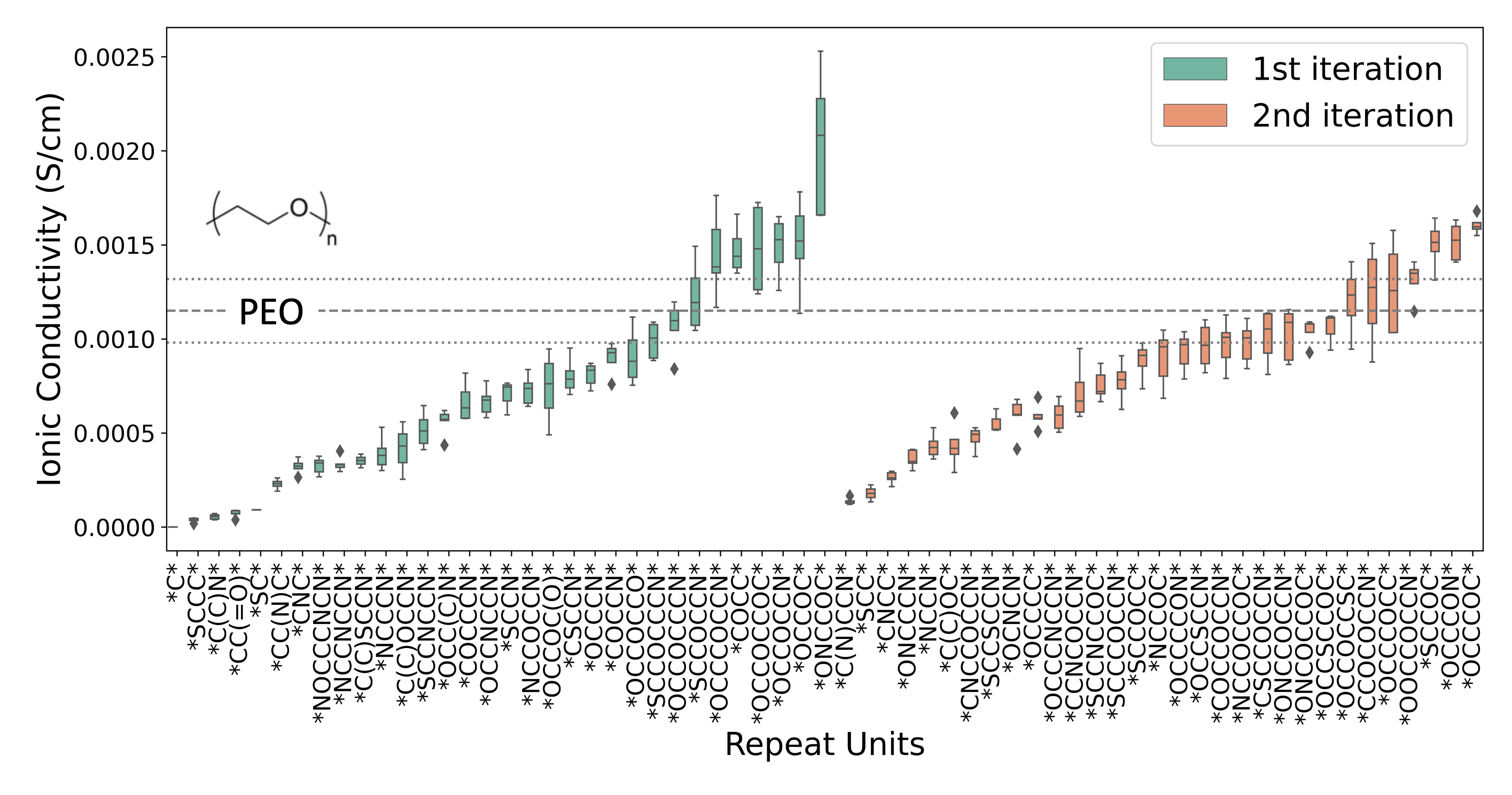}
    \caption{Ionic conductivity of polymers generated from two iterative candidate generations computed from MD simulations. The box plots show the mean and standard deviation in 5 MD simulations performed for each listed polymer. The dashed and dotted lines show the mean and the standard deviation of ionic conductivity of PEO computed from MD simulation.}
    \label{fig:self-improvement}
\end{figure}

\begin{figure}[h]
    \centering
    \includegraphics[width=0.9\textwidth]{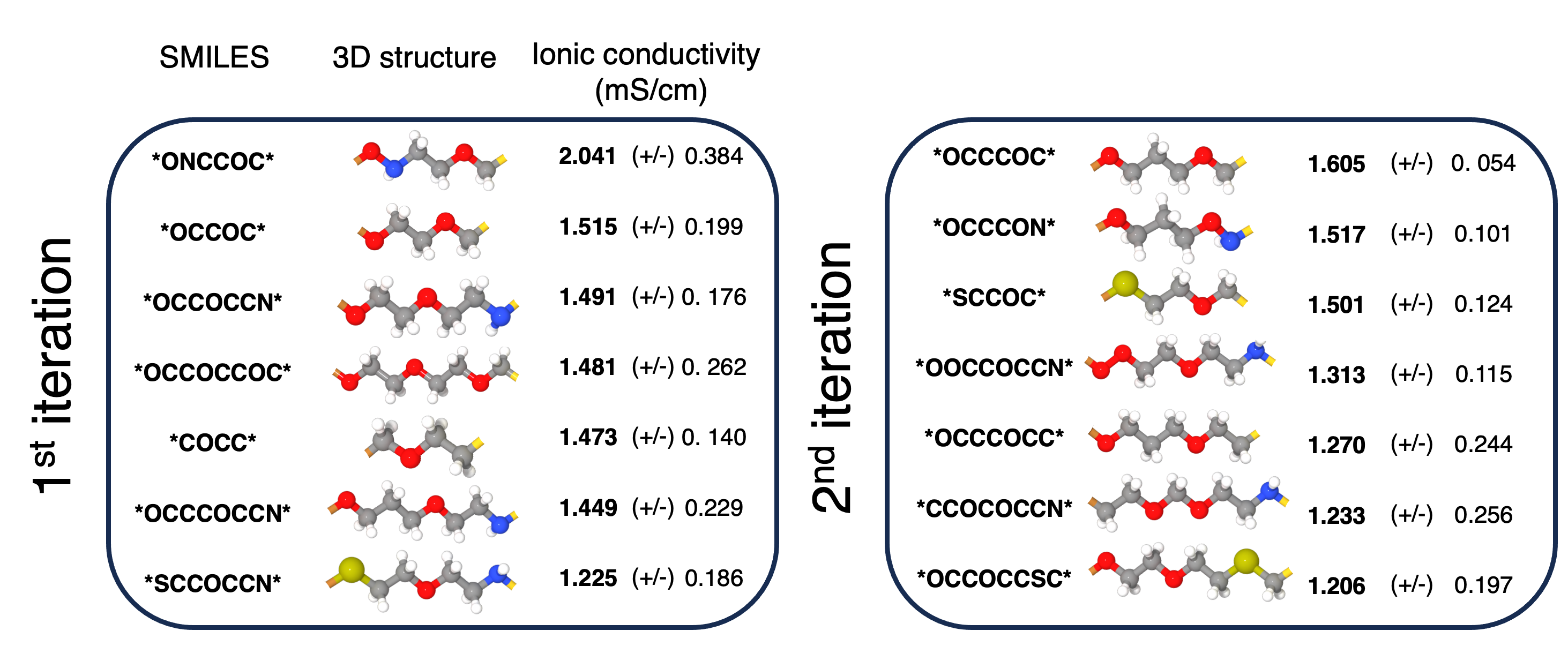}
    \caption{Discovered polymers from two iterative generation cycles. The polymer listed for each iteration exhibited an ionic conductivity superior to that of PEO.}
    \label{fig:discovered-polymers}
\end{figure}

The calculated ionic conductivity, derived from the cluster Nernst-Einstein equation, arises from both ion diffusivity and clustering. To elucidate the mechanisms underpinning the superior ionic conductivity observed in generated polymers, we conducted a comparative analysis of conductivity, ion diffusivity, and concentrations of free ions between PEO and the polymers exhibiting enhanced performance (see Fig.S1). In this context, "free ions" denote those not incorporated into any clusters and moving freely, with their concentration determined as an average across simulation durations. The analysis reveals that the augmented conductivity in the most effective polymers from the first two iterations is attributed to both an increase in ion diffusivity and a higher prevalence of free ion clusters. Interestingly, it was also noted that several of the developed polymers exhibit a more efficient dissociation of the $Li^+.TFSI^-$ salt compared to PEO, indicating a potential for improved ion transport properties.

Salt concentration is a crucial design parameter that influences ion pairing in polymer electrolytes. Both ionic conductivity and free ion concentration initially increase with higher salt concentration, but at very high concentrations, ion clusters of cations and anions form. These clusters reduce effective conductivity due to charge cancellation within each cluster~\cite{PhysRevLett.122.136001}. The optimal salt concentration depends on how strongly polymer atoms coordinate with ions, which affects salt dissociation. Consequently, this optimal concentration varies for different polymers. Generally, a practical electrolyte system for lithium-ion batteries requires a moderate to high salt concentration to achieve enhanced ionic conductivity~\cite{doi:10.1021/jp2111956}, mechanical strength~\cite{mohanty2022effect}, electrochemical stability~\cite{https://doi.org/10.1002/polb.24235}, and solid electrolyte interphase (SEI) formation~\cite{yoon2017lithium}. We examined the ionic conductivity of several top candidates across various salt concentrations at 353 K to illustrate this concept (see Fig. S2). The MD simulation results indicated that maximum conductivity occurs at slightly different salt molalities, generally around 1.5-2.0 mol/kg. Although the salt concentration in the training set used to generate the polymers in this study was 1.5 mol/kg, exploring this parameter further as a design factor is recommended for future research.

\section{Discussion and Conclusion}
As we reflect on the undeniable capabilities of our discovery platform, it is equally critical to acknowledge its current limitations and the vast potential for future enhancements. This recognition not only grounds our work in a realistic context but also opens avenues for exciting and impactful future research.

As mentioned earlier, the conditioned generative model, in its present state, is somewhat crude and unrefined. The model's efficacy as the primary driver of polymer discovery underscores the need for further development and refinement. Although the choice of using SMILES strings to represent materials allows us to apply the presented discovery framework to other cases, such as molecular discovery (e.g., liquid electrolytes), it also limits the information available to the model. SMILES strings are designed to represent molecules as linear sequences, which can be inadequate for capturing the complexity of branched or cross-linked polymer structures. However, it is possible to enhance SMILES strings with 3D structural information or to develop representations that are inherently aware of 3D structures. Data oversampling strategies that have proven to be instrumental in the experiment also need to be thoroughly and systematically investigated. Future efforts could focus on improving the current model or exploring alternative architectures, potentially enhancing the platform's efficiency and accuracy in generating high-quality polymer candidates. Notably, developing generative algorithms for such specialized tasks presents its own set of challenges, particularly in evaluating their effectiveness. Unlike more straightforward metrics such as Mean Absolute Error (MAE) used in regression tasks, assessing the performance of generative models is less direct. A sister publication~\cite{Yang2023}, which is released concurrently, delves deep into the benchmark of different generative model architectures in various generative tasks. It addresses the issue of evaluation by proposing a comprehensive and systematic methodology to benchmark different models for polymer SMILES code generation. This approach will significantly aid in comparing and optimizing generative models for polymer discovery.

In this study, we selected ionic conductivity as the primary metric to identify new polymer electrolytes. However, as highlighted in previous studies~\cite{snyder2021improved, halat2021modifying}, a more holistic measure is efficacy, defined as the product of conductivity and cationic current fraction. This measure considers both conductivity and the mobility of the preferred charge (cation here) in the system. In our study, ion conductivity calculations were standardized at a salt concentration of 1.5 mol/kg, which is around the optimum for ion mobility for PEO. Moving forward, we plan to modify our evaluation criteria to include ion transport efficacy based on cation mobility and to establish a database encompassing various salt concentrations. 

Additionally, to address the growing demands for multi-properties optimization for materials science problems, our platform can be adapted to evaluate crucial polymer properties such as glass transition temperature (Tg), bulk density, and mechanical properties. This would involve extending the current framework to incorporate these properties into the design criteria, allowing for a more comprehensive assessment of polymer candidates. Future research will explore incorporating these additional properties into the model input to enhance the platform’s capability in identifying polymers that not only exhibit superior ionic conductivity but also meet other critical performance metrics. We would also like to emphasize that beyond properties such as ionic conductivity, other critical factors are essential for successful polymer discovery, with synthesizability being one of the most important. Our platform can indeed incorporate considerations of synthesizability. For tools that quantify synthesizability, such as the SA score, we can generate and iterate on more synthesizable candidates in the same manner as we did for ion conductivity. For retrosynthesis planning tools~\cite{chen2021data}, which propose synthesis routes for candidates, we can a) quantify synthesizability based on factors such as the number of plausible synthesis recipes, required synthesis conditions, and the kinetics of the routes, and use that as a condition; and/or b) enhance the feedback loop by incorporating real-world testing of the synthesis recipes.

Another vital aspect of our platform is the computational evaluation process. Currently, we employ classical molecular dynamics simulations, which are fast and effective but may lack robustness. The ultimate goal is to corroborate our findings through experimental validation. However, given the current limitations in high-throughput polymer synthesis and characterization, simulation-based assessment remains the most feasible approach. Our future endeavors will likely involve evaluating candidates through simulations before proceeding with experimental trials. Furthermore, the reliance on classical force fields, which are typically tailored to specific polymer classes, raises concerns about their extrapolative power to other materials. There is a noticeable disagreement between the reported ionic conductivity for the poly(EO-MO) (OCCOC) from ref\cite{snyder2021improved, halat2021modifying} and our computed one, which could be a result of the artifact of the force field. To address this, one promising direction is the use of ab initio molecular dynamics (AIMD) simulations with methods like Density Functional Theory (DFT). While these simulations offer greater robustness, their computational expense is a significant barrier. A potential solution could be the development and use of machine-learned force fields that combine the accuracy of DFT with the efficiency of classical force fields. Additionally, experimental verification would be critical for the promising candidates.

Although the workflow of our platform currently lacks full modularity and automation to provide more flexibility and expedite development, this approach presents an opportunity for further improvement. Throughout the discovery campaign, the process included minimal human intervention, laying the groundwork for developing a fully automated system. Future efforts will be directed toward enhancing the platform's modularity, enabling compatibility with various generative models and evaluation methods. This improvement would not only streamline the discovery process but also expand the platform's utility across different domains of polymer research. 

\section{Code and Data Availability}
The dataset used to train the generative models can be accessed on the HTP-MD website: \href{https://www.htpmd.matr.io/}{https://www.htpmd.matr.io/}. The subset of HTP-MD dataset used for training has been included in the \href{https://github.com/TRI-AMDD/PolyGen}{https://github.com/TRI-AMDD/PolyGen} as {htpmd-trainset.csv}.   The code for training the generative models can be found here: \href{https://github.com/TRI-AMDD/PolyGen}{https://github.com/TRI-AMDD/PolyGen}. The code for running the MD simulations can be found here: \href{https://github.com/TRI-AMDD/PolyGen/tree/main/Example-simulation-files}{https://github.com/TRI-AMDD/PolyGen/tree/main/Example-simulation-files}. Consistent with the trainset, the MD simulation trajectories were analyzed, and the ionic conductivity of the generated polymers has been computed using HTP-MD code available at \href{https://github.com/TRI-AMDD/htp_md}{https://github.com/TRI-AMDD/htp$\_$md}. 

\section{Disclosures}

The authors wish to acknowledge that the discovery framework and materials discovered using our generative model framework, as described within this manuscript, are subject to a provisional patent application. This application has been submitted with the following details: U.S. Patent Application No. 63/582,871 titled "Methods of Designing Polymers and Polymers Designed Therefrom" with TEMA Reference No. IP-A-6823PROV and Darrow Reference No. TRI-1107-PR. The authors confirm that this does not alter our adherence to ArXiv policies on sharing data and materials.

\section*{Acknowledgments}
We thank Professors Grossman, Yang Shao-Horn, Jeremiah Johnson, and Rafael Gomez Bombarelli, and Drs. Tian Xie, Sheng Gong, and Arthur France-Lanord at the Massachusetts Institute of Technology for their critical support in our molecular dynamics simulations and polymer electrolyte research. Their guidance and insightful discussions have greatly enhanced our study's development and robustness.


\bibliographystyle{unsrt}  
\bibliography{references}  

\end{document}


\title{Supplementary Information - A Self-Improvable Polymer Discovery Framework Based on Conditional Generative Model}
\maketitle

\centering Khajeh et al.

\begin{table}[h]
    \centering
    \caption{Hyperparameter selection for the GPT model.~\citep{Yang2023}}
    \includegraphics[width=1.0\textwidth]{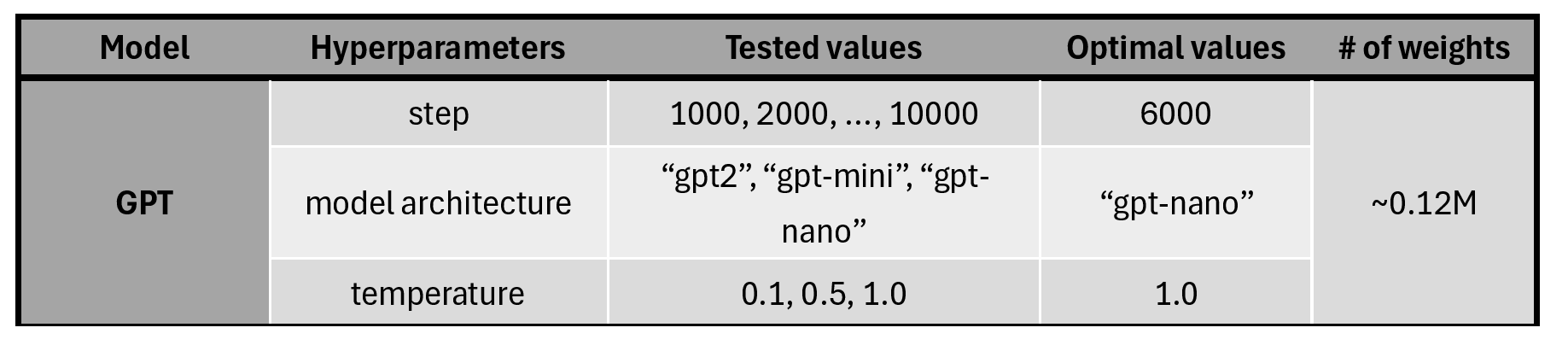}
    \label{tab:hyperparameters_selection}
\end{table}

\begin{table}[h]
    \centering
    \caption{Grid search results for the GPT model (part 1).~\citep{Yang2023}}
    \includegraphics[width=1.0\textwidth]{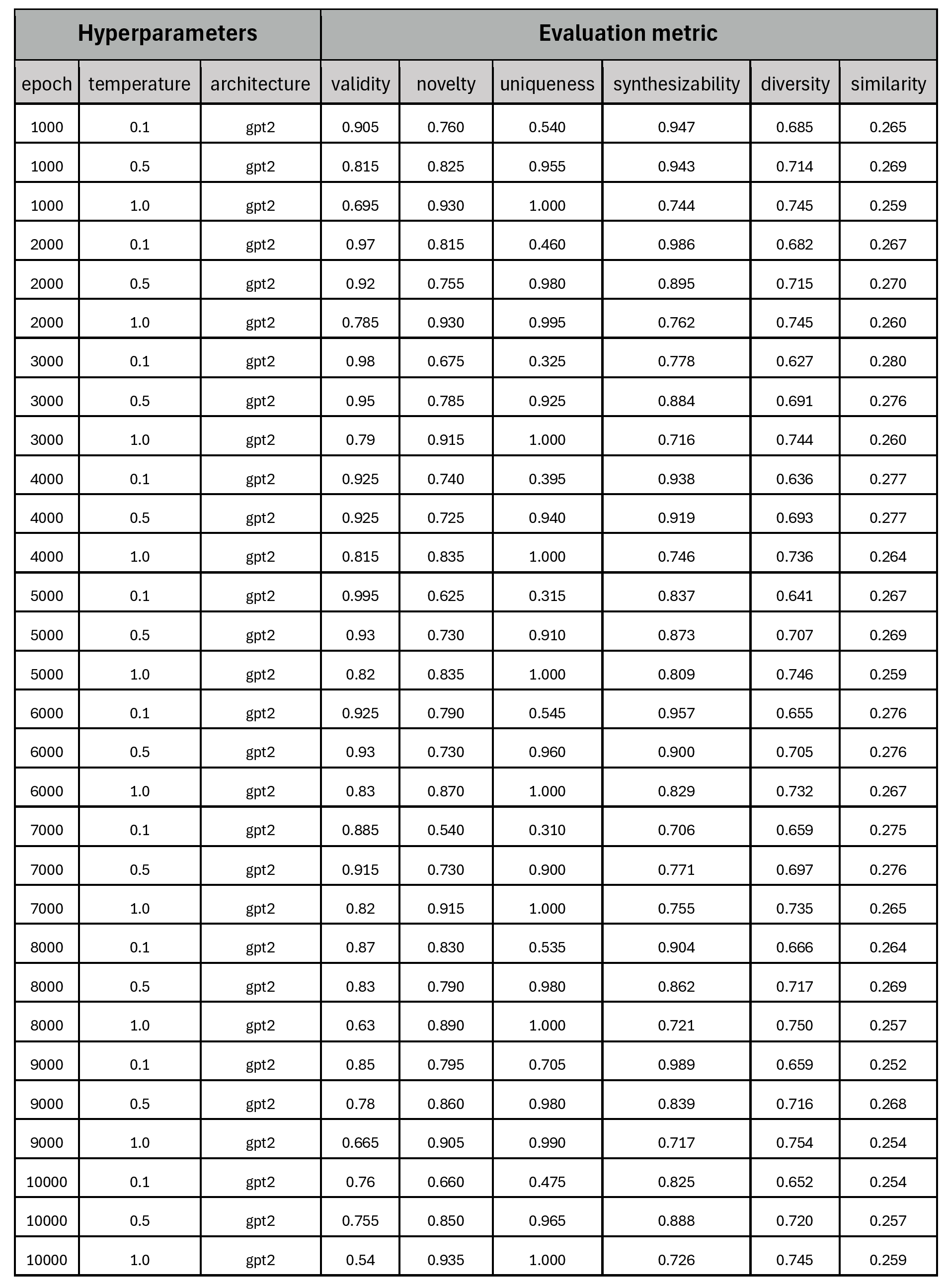}
    \label{tab:grid-search-1}
\end{table}

\begin{table}[h]
    \centering
    \caption{Grid search results for the GPT model (part 2).~\citep{Yang2023}}
    \includegraphics[width=1.0\textwidth]{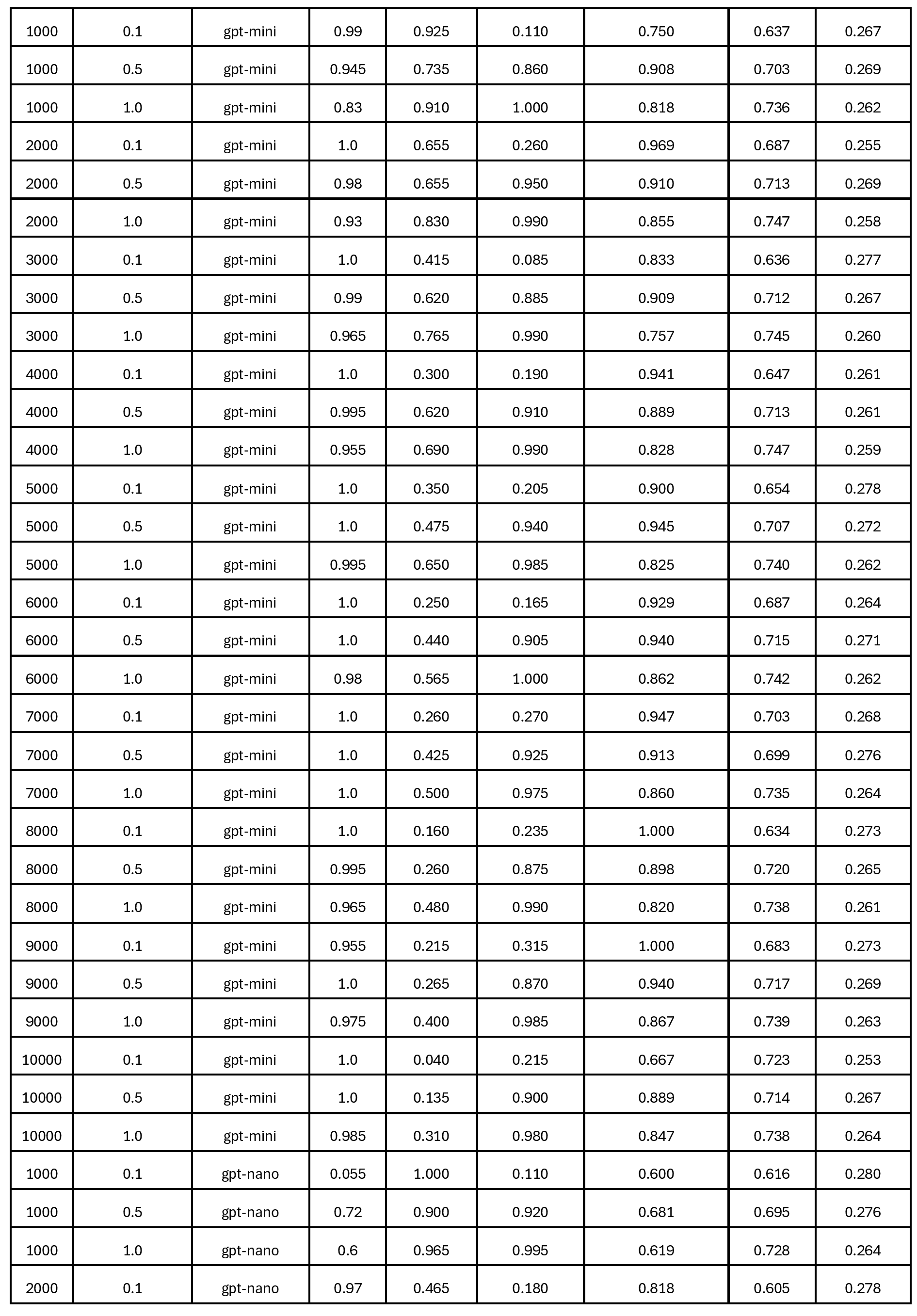}
    \label{tab:grid-search-2}
\end{table}

\begin{table}[h]
    \centering
    \caption{Grid search results for the GPT model (part 3).~\citep{Yang2023}}
    \includegraphics[width=1.0\textwidth]{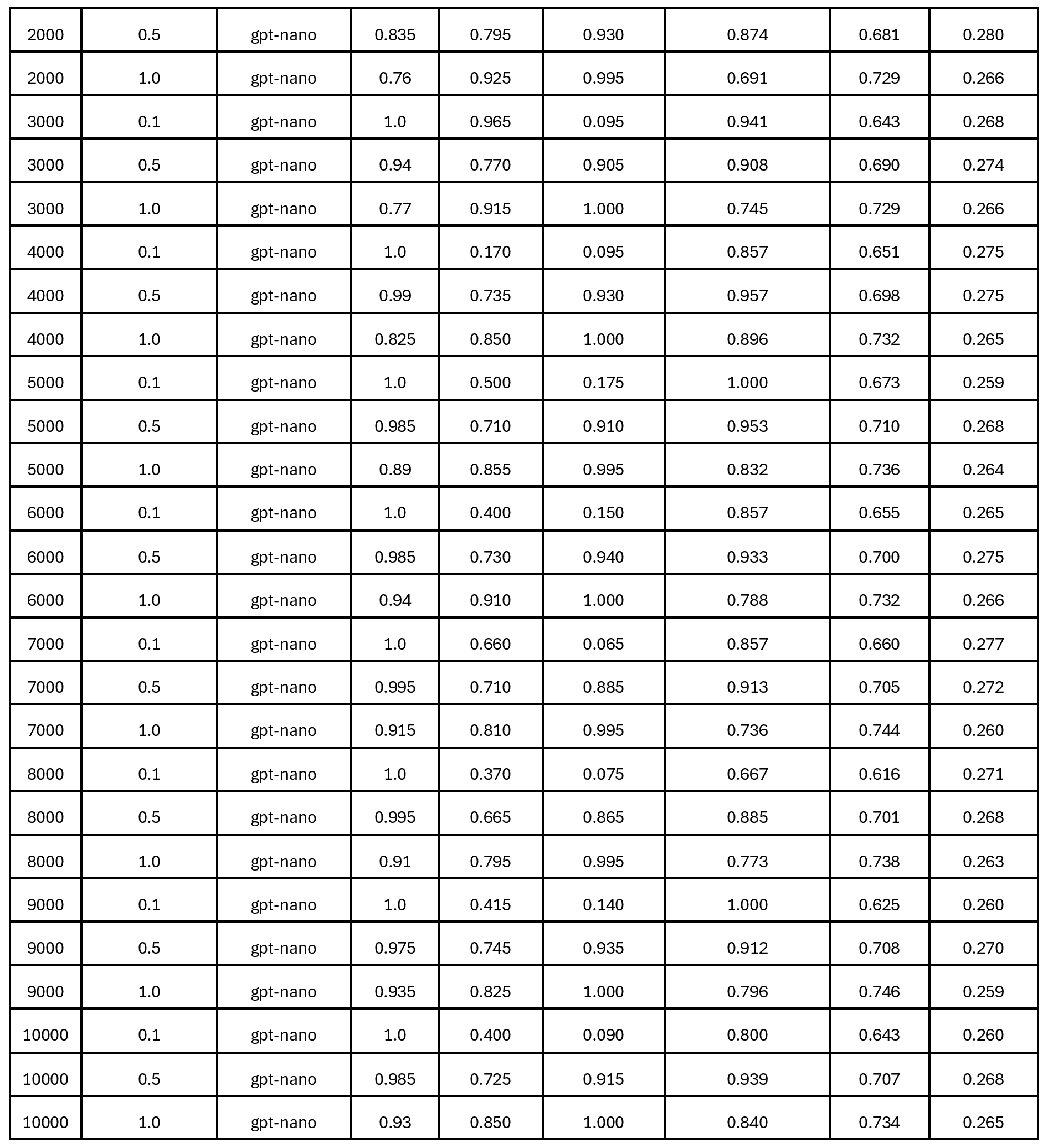}
    \label{tab:grid-search-3}
\end{table}

\begin{table}[h]
    \centering
    \caption{Average of computed ion transport properties of generated polymers in the 1$^st$ iteration}
    \includegraphics[width=1.0\textwidth]{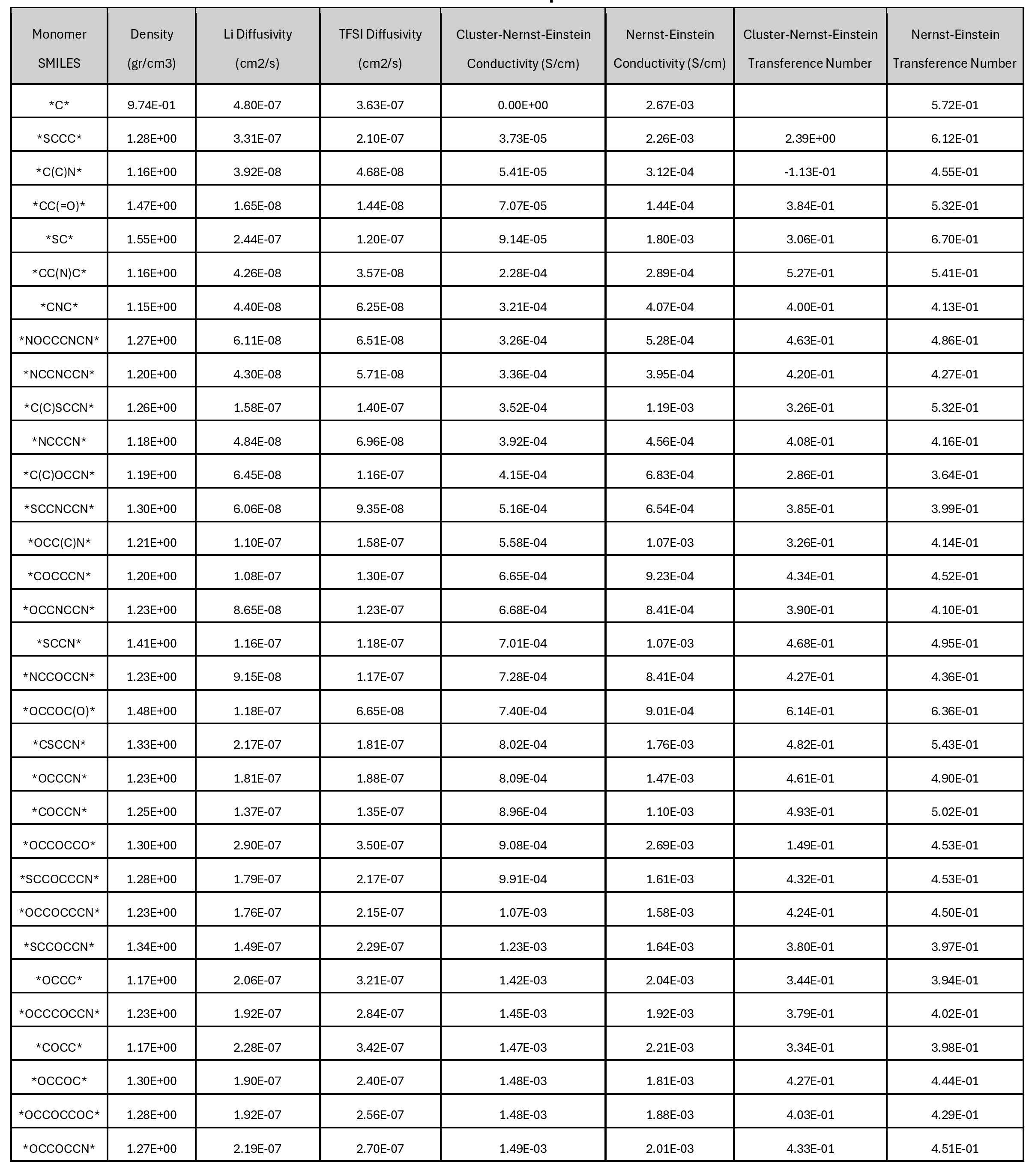}
    \label{tab:iteration-1}
\end{table}

\begin{table}[h]
    \centering
    \caption{Average of computed ion transport properties of generated polymers in the 2$^nd$ Iteration}
    \includegraphics[width=1.0\textwidth]{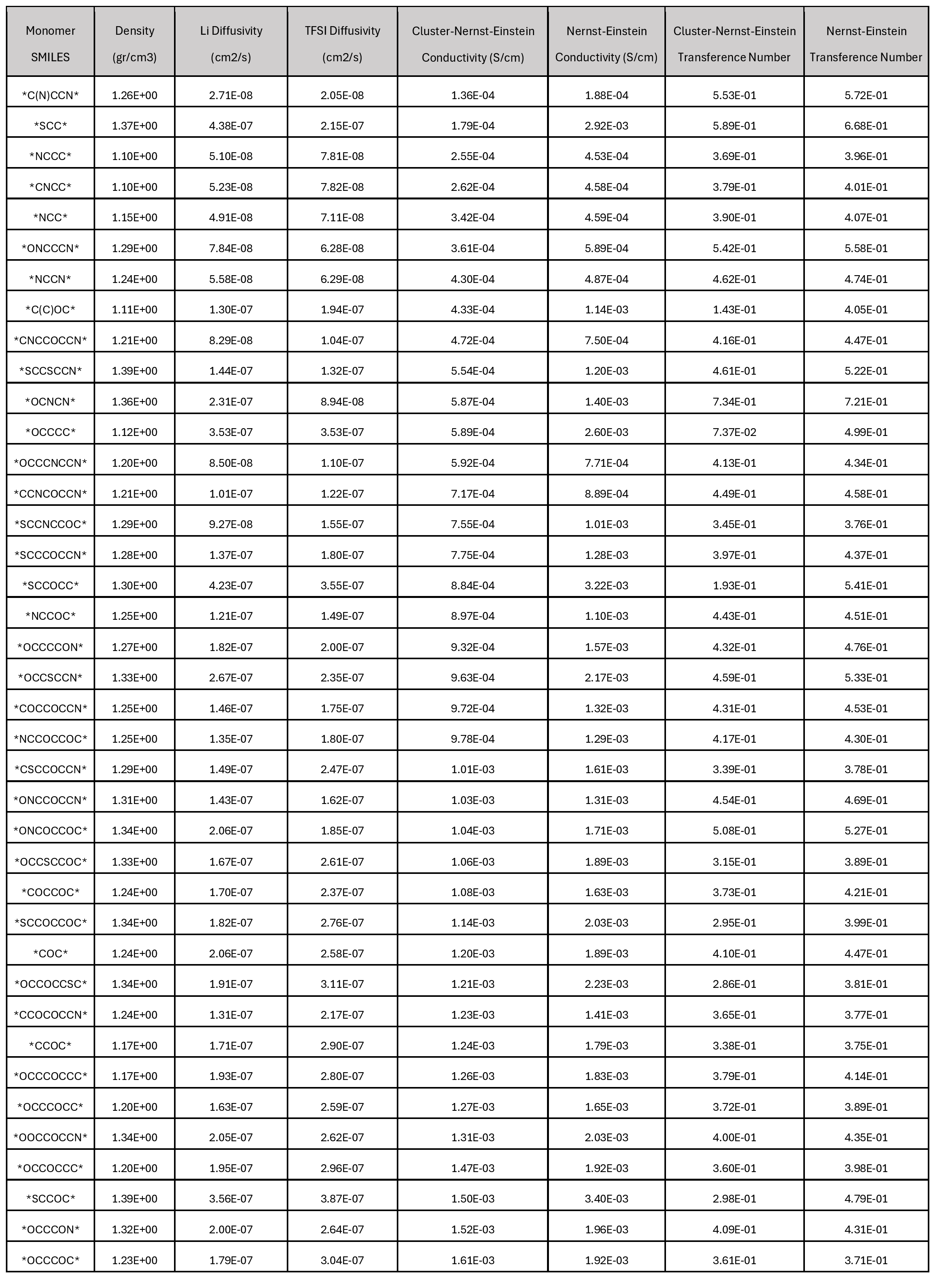}
    \label{fig:iteration-2}
\end{table}

\begin{figure}[h]
    \centering
    \includegraphics[width=0.9\textwidth]{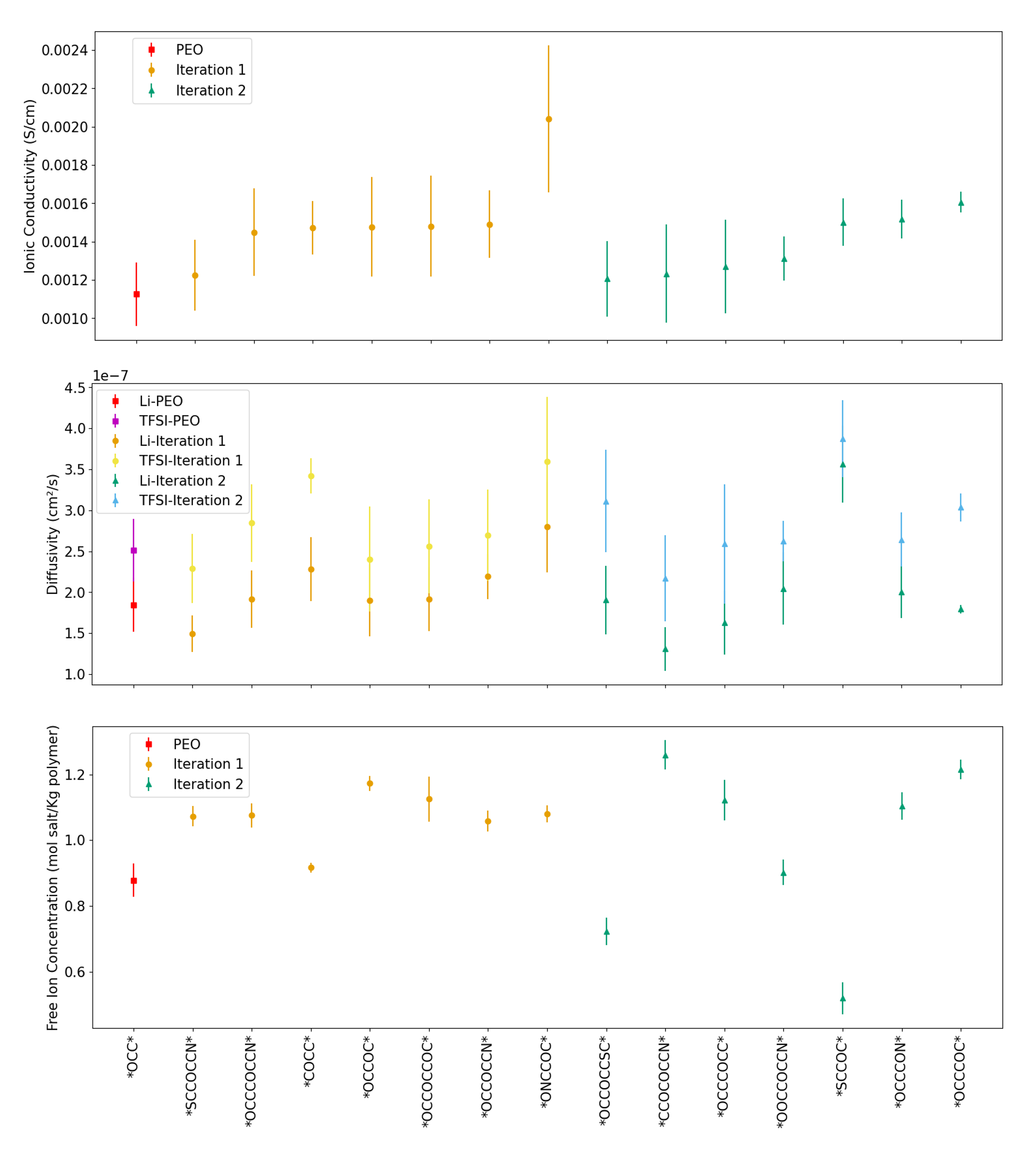}
    \caption{Comparison of the ionic conductivity, ions' diffusivity, and free ion concentration between PEO and the generated polymers with higher conductivity. The first point in each plot represents the ion transport properties of PEO. All reported data in this graph are computed from molecular dynamics (MD) simulations.}
    \label{fig:origins-of-high-conductivity}
\end{figure}

\begin{figure}[h]
    \centering
    \includegraphics[width=0.9\textwidth]{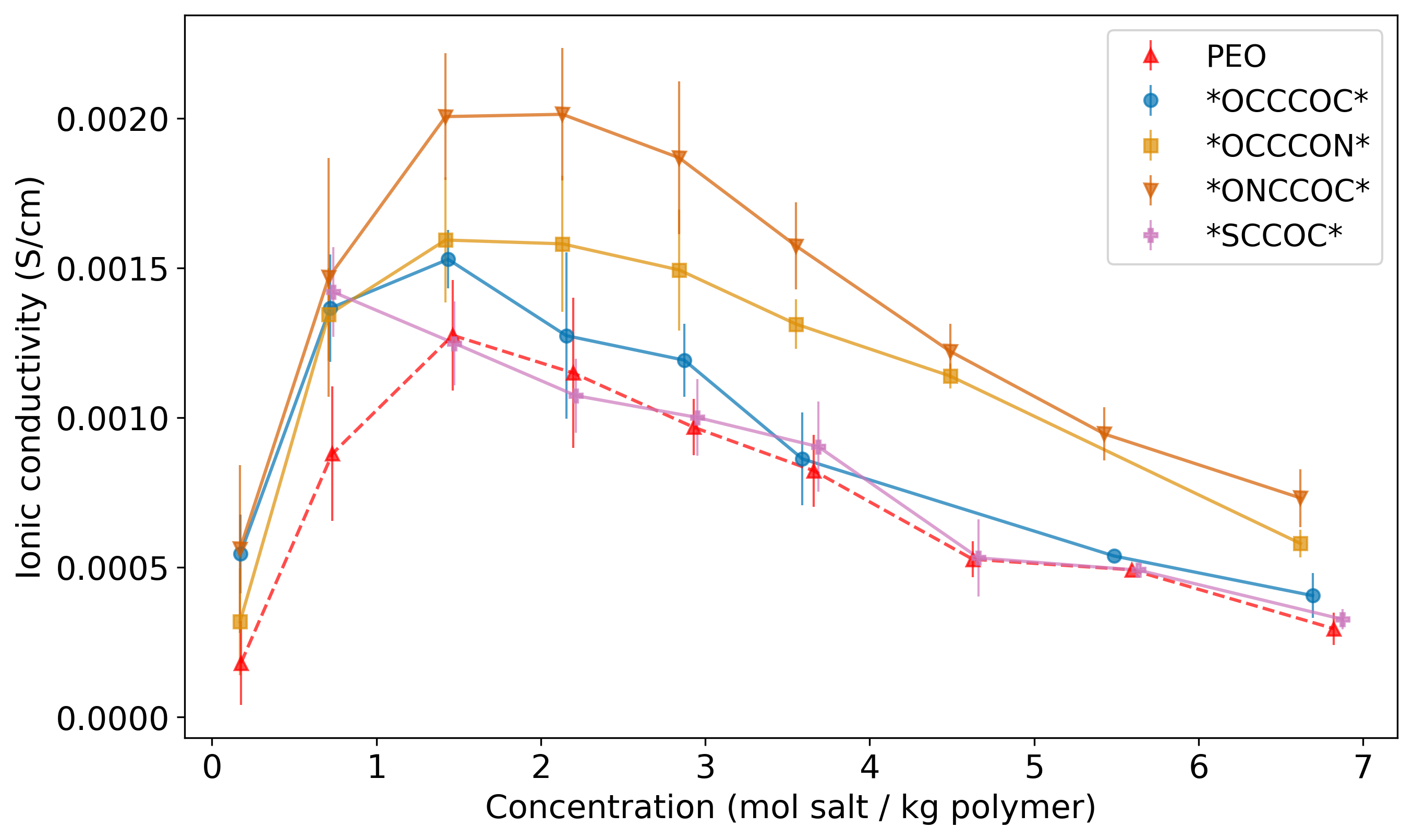}
    \caption{Effect of salt concentration on ionic conductivity captured in MD simulations. The position of maximum ionic conductivity depends on the polymer structure and occurs at different salt molalities for different polymers.}
    \label{fig:effect-of-salt-concentration}
\end{figure}

\clearpage
\bibliographystyle{plain} 
\bibliography{references}